\documentclass[
 amsmath,amssymb,
 aps,
prb,
floatfix,
twocolumn, 10pt
]{revtex4-1}

\usepackage{CJKutf8}
\usepackage{graphicx}% Include figure files
\usepackage{bm}% bold math
\usepackage{wasysym}
\usepackage{bbold}
\usepackage{hyperref}% add hypertext capabilities

\begin{document}

\title{Kramers' revenge}

\author{Pengke Li (\begin{CJK*}{UTF8}{gbsn}李鹏科\end{CJK*})}
\email{pengke@umd.edu}
\author{Ian Appelbaum}
\email{appelbaum@physics.umd.edu}
\affiliation{Department of Physics,
U. of Maryland, College Park, MD 20742}

\begin{abstract}
The combination of space inversion and time reversal symmetries results in doubly-degenerate Bloch states with opposite spin. Many lattices with these symmetries can be constructed by combining a noncentrosymmetric potential (lacking this degeneracy) with its inverted copy. Using simple models, we unravel the evolution of local spin-splitting during this process of inversion symmetry restoration, in the presence of spin-orbit interaction and sublattice coupling. Importantly, through an analysis of quantum mechanical commutativity, we examine the difficulty of identifying states that are simultaneously spatially segregated and spin polarized. We also explain how surface-sensitive experimental probes (such as angle-resolved photoemission spectroscopy, or ARPES) of `hidden spin polarization' in layered materials are susceptible to unrelated spin splitting intrinsically induced by broken inversion symmetry at the surface.
\end{abstract}

\maketitle

\section{Introduction}
In crystalline electronic materials, the energy dispersion relation for a given band, $E_\sigma(\bm{k})$ (where $\sigma$ indicates spin up or down and $\bm{k}$ is quasimomentum), is constrained by the exact symmetries of the underlying Hamiltonian. Time-reversal symmetry ($\mathcal{T}$) switches both $\bm{k}\rightarrow-\bm{k}$ and $\sigma\rightarrow -\sigma$, leading to $E_\uparrow(\bm{k})=E_\downarrow(-\bm{k})$\cite{Kramers_ProcA30, Wigner_NWG32, Bouckaert_PR36, Herring_PR37}. Space inversion symmetry $\mathcal{I}$, on the other hand, only transforms $\bm{k}\rightarrow-\bm{k}$, giving $E_\uparrow(\bm{k})=E_\uparrow(-\bm{k})$. The presence of both symmetries protects the double degeneracy $E_\uparrow(\bm{k})= E_\downarrow(\bm{k})$ of states at arbitrary $\bm{k}$ with opposite spins\cite{Elliott_PR54b, Dresselhaus_book}. In lattices that lack inversion symmetry, such degeneracy is generally broken by the spin-orbit interaction (SOI), resulting in so-called `spin-splitting', except at time-reversal-invariant momentum (TRIM) points, and along certain axes of high spatial symmetry having double group irreducible representations of even dimension \cite{Cardona_PRB88}.

Inversion-asymmetric {\em two dimensional} lattices (such as transition-metal dichalcogenides\cite{Song_PRL13}, three-six-enes\cite{Li_PRB15} and four-six-enes\cite{Appelbaum_PRB16}) provide an interesting building block to explore the effect of inversion parity, because appropriate stacking of even numbers of layers can result in a larger lattice that restores inversion symmetry. However, one must be cautious with assumptions such as ``solids with spatial inversion symmetry do not display spin–orbit effects'' \cite{Partoens_NatPhys14}. When two complementary layers (both noncentrosymmetric) are separated by an arbitrarily large distance or coupled by vanishingly weak interactions, inspection of a virtually isolated single layer will still presumably reveal local spin splitting at general wavevector $\bm{k}$, even though the Bloch band double degeneracy of the entire centrosymmetric system must be preserved.

This apparent conundrum can be resolved by distinguishing between the double degeneracy caused by the combination of $\mathcal{T}$ and $\mathcal{I}$, and `pure' spin degeneracy, where two states with the same orbital wavefunction have opposite spins. The former case pairs two states whose spatial parts transform into the other's complex conjugate via inversion, with orthogonal spin due to time-reversal. Importantly, this inversion pairing does not necessarily map a given spatial wavefunction back onto itself, as seen in the aforementioned example of inversion-related 2D layers. Indeed, in many 3D centrosymmetric lattices, the inversion operation does not leave each lattice site invariant (with the assistance of periodic translations from the full space group), so the lattice can be decomposed into complementary `sectors' paired by inversion. One notable example is the dual (and also centrosymmetric) face-centered cubic (FCC) sublattices together forming a diamond structure. 

Recently, this type of composite lattice was studied using density functional calculations, in which the projection of the wavefunctions onto each of the paired `sectors' was found to be `spin-polarized'\cite{Zhang_NatPhys14}. Later, the same method was adapted to a number of layered two-dimensional van~der~Waals lattices\cite{Liu_PRB15}, where the equal but opposite (and hence fully compensating) band spin splitting of the two sublattices was attributed to the locally noncentrosymmetric point-group symmetry.

Partly due to its computational complexity, density functional schemes may obscure the underlying physics of nominally complicated phenomena, causing confusion. In this case, some were inspired to suggest rewriting textbooks on the elemental pillars of the subject \cite{Partoens_NatPhys14}. Experimental efforts employing Angle-Resolved Photoemission Spectroscopy (ARPES) have been similarly motivated to make empirical contact to the computational results\cite{Riley_NatPhys14,Bertoni_PRL16, Cottin_NatComm16, Razzoli_PRL17, Yao_NComm17}.

One way to clarify and qualify the meaning of this `hidden spin polarization' used in the context of complicated numerical computation is to analyze the system from a diametric viewpoint. Here, we use simple noncentrosymmetric models as complementary `sectors' to rebuild a globally centrosymmetric system, analytically revealing the evolution of local `spin-polarization' during restoration of inversion symmetry. We will: 
(1.) show that Hamiltonian commutativity with a unique spin operator is absent, spoiling spin purity of spatially segregated states, 
(2.) explain that any experimental measurement designed to corroborate sector-dependent `spin polarization' requires explicit inversion symmetry breaking, and 
(3.) show how surface-sensitive electronic structure analysis methods such as ARPES entangle spin splitting caused by symmetry-reducing truncation of the lattice with the sublattice-projected `spin polarization' sought.

\section{\textit{Mise en place}: preliminaries}
`Spin orbit interaction' (SOI) $ \frac{\hbar}{4m^2c^2}\nabla V\times \bm {p} \cdot \bm{\sigma}$ is an essential ingredient in the physics of spin splitting within electronic structure. It formally arises as a term in the electron Hamiltonian from a leading-order expansion of the Dirac equation, and can be viewed as an inescapable result of electromagnetic Lorentz invariance. We distinguish the SOI operator from its consequences in a given spectrum by calling the latter `spin-orbit {\textit{coupling}}' (SOC). SOC manifests in different ways, depending on the bands' orbital symmetry that can be classified using representation theory\cite{Bouckaert_PR36, Herring_PR37, Elliott_PR54b, Lok_book, Cardona_PRB88}. For example, SOC in the valence band of cubic lattices appears as broken threefold orbital degeneracy of ($\ell=1$) $\Gamma_4$ bands at $\bm{k}=0$ regardless of inversion symmetry, resulting in the well-known ($j=1/2$) $\Gamma_7$ `split-off' band, separated from the remaining fourfold-degenerate ($j=3/2$) $\Gamma_8$  light- and heavy-hole band by an amount of energy that is dependent on the details of electric fields near the atomic core; in general, higher atomic numbers result in larger splitting.

Band symmetry also plays a crucial role in determining the SOC effects away from TRIM points. A familiar example is given by zincblende, where the lowest-order momentum-dependent (Dresselhaus) spin splitting is cubic in $k$ for the $\Gamma_6$ conduction band, and linear for the $\Gamma_{7,8}$ valence bands, except along lines of high symmetry where it vanishes\cite{Dresselhaus_PR55, Cardona_PRB88, Lok_book}. Such spin splittings originate from the absence of an inversion center in the lattice; structural configurations lacking inversion symmetry can also cause spin splitting. One well-known illustration is given by the two dimensional electron gas bound to a single heterostructure interface, where (Bychkov-Rashba) spin splitting depends on the potential gradient normal to the interface and is linear in $k$ to lowest order\cite{Bychkov_JETPL84}.

From a semiclassical viewpoint, these $\bm k$-dependent spin splittings can be understood to result from an effective Zeeman interaction with the momentum-dependent magnetic field $\propto\bm{\nabla} V\times \bm{\hat{p}}$.  Whereas $\bm{\nabla} V$ always averages to zero in any periodic lattice, its cross product with momentum operator $\bm \hat{p}$ is nonzero, {\textit{except}} in centrosymmetric potentials. However, this does not mean that SOC is somehow absent when inversion symmetry is present: it still manifests as broken orbital degeneracy (see above) and as an intermixing of wavefunctions with opposite spin from remote bands. SOC is therefore the latent cause of scattering-driven spin flip responsible for finite spin lifetime at nonzero temperatures\cite{Elliott_PR54,Yafet_SSP63}, and has motivated the study of various spin transport phenomena in centrosymmetric materials for many years\cite{Guo_PRL07, Huang_PRL07, Han_PRL11, Li_PRL13}.

Care must be taken in using the phrase `spin polarization'. First of all, there is zero {\em ensemble} spin imbalance of occupied states in equilibrium, as guaranteed by Kramers' theorem. However, when applied to a single-particle spectrum, the phrase `spin polarization' requires that there exists a \textit{unique} spin operator $\hat{S}$ which \textit{commutes} with the underlying Hamiltonian, so that energy eigenstates can also simultaneously be (mutually orthogonal) spin eigenstates.  Projecting out components of the wavefunction residing solely on one sublattice with projection operator $\hat{P}$ as in Ref.~[\onlinecite{Zhang_NatPhys14}] transforms our spin operator into $\hat{P}^\dagger\hat{S}\hat{P}$, which in general no longer commutes with the Hamiltonian. Its expectation values do not represent a bona fide stationary property endowed by the full Hamiltonian. To some limited extent, they do reflect properties of observables of the  \textit{degenerately} perturbed `sector', but only under weak coupling conditions when coupling energy is small with respect to spin splitting in a single sublattice. Moreover, in the presence of double degeneracy associated with inversion symmetry, the decomposition of the lattice Hamiltonian into two complementary `sectors' is arbitrary up to a unitary transformation, so the resulting expectation value of $\hat{P}^\dagger\hat{S}\hat{P}$ depends on the choice of `sectors' and is not gauge invariant. 

More generally, the Hamiltonians of real systems are usually modeled within a Hilbert space spanned by more than just two basis functions in the spatial subspace. For example, a tight-binding model of valence electrons could include $s$, $p$, $d$ and even further atomic orbitals\cite{Jancu_PRB98}, whereas the orthogonal planewave formalism considers a basis consisting of many reciprocal lattice vectors\cite{Chelikowsky_PRB76}. One can project out a single basis component of a given eigenstate, but expectation values of the spin operator (or other quantum mechanical operators associated with physical observables) in this purified Hilbert space will necessarily vary depending on the choice of basis. 

The physical meaning behind such contrivances is questionable without experimental capability to resolve spatial, orbital, or Fourier components of the wavefunctions. Although `hidden spin polarization' at first appears to be a surprising discovery in systems possessing space inversion symmetry, it is derived from projection of eigenstates onto the `sector' subspace, so any possible experimental detection requires inequivalence between the two complementary sectors of the delocalized wavefunctions. It therefore \textit{explicitly requires inversion symmetry breaking}, under which spin-splitting is definitely not unusual. As a surface sensitive technique that can resolve the band structure in reciprocal space, ARPES naturally provides a projection operation to select the top surface `sector' of wavefunctions in  bilayer systems or very thin films. However, for samples thicker than several atomic layers, truly bulk states are beyond the detection capability of ARPES,\cite{Armitage_RMP18} regardless if their centrosymmetric lattices can be decomposed into complementary `sectors'. As we will show, even at a perfect surface of the centrosymmetric simple cubic lattice (which cannot be decomposed into complementary `sectors' unrelated by full lattice translations), the surface band structure is inevitably spin-split by an effective electric field that otherwise vanishes in the bulk. This generic source of spin splitting at an arbitrary surface interferes with any splitting due to the noncentrosymmetry of the exposed `sector' targeted by ARPES.

\section{Toy models}

As mentioned previously, the band structure of any inversion asymmetric system is spin-split by SOI at general $\bm k$.  In this section, we use two simple models (1D chains and 2D planes) to show in a pedagogical way how the spin-dependent eigenstates evolve when the double degeneracy is restored by compensating this splitting with a spatially inverted copy. At the end, we comment on `sectors' with inversion symmetry.

\subsection{Tight-binding}

\begin{figure}[t!]
\includegraphics[width=3.3in]{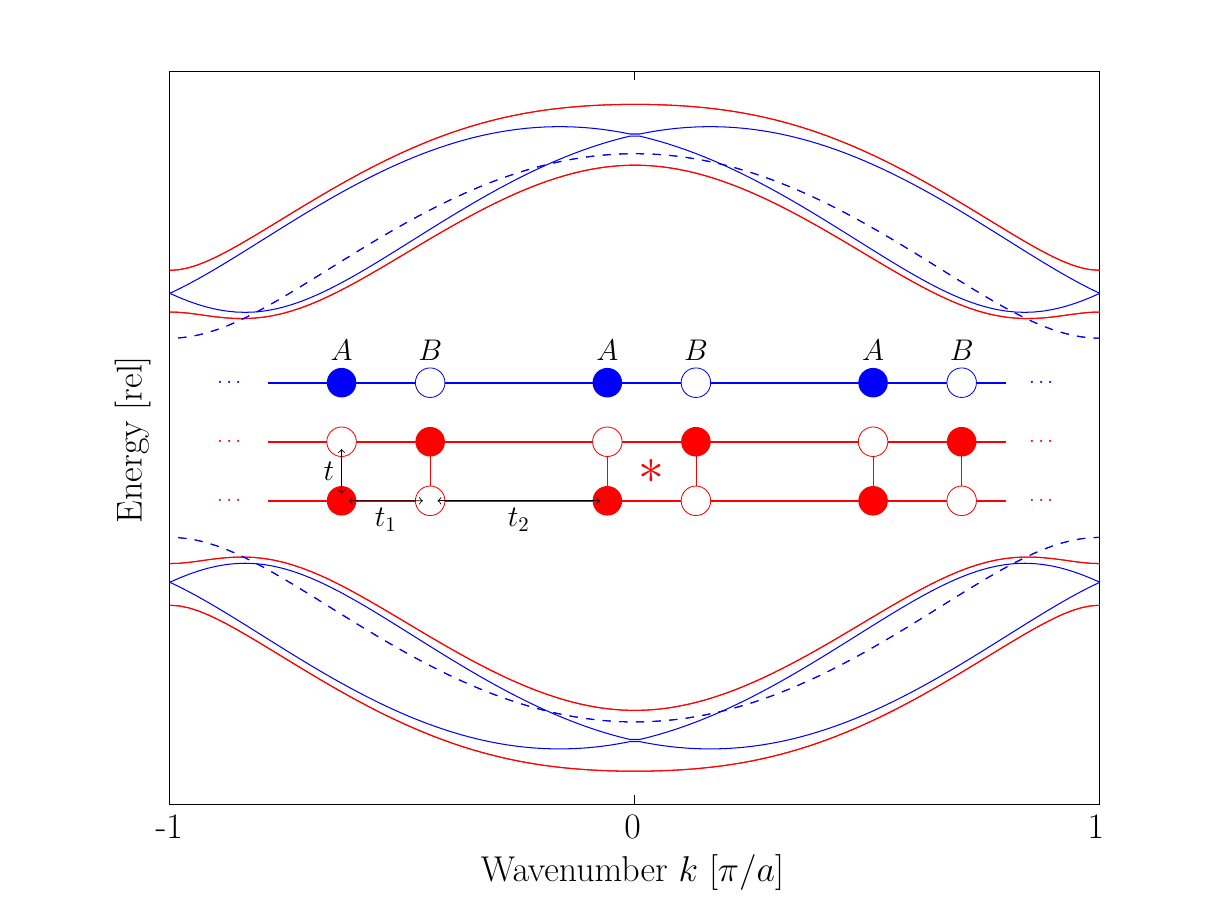}
\caption{Tight-binding bandstructure of a single 1-D inversion asymmetric chain without SOC (dashed blue, Eq. \ref{eq:E0}) and with SOC (solid blue, Eq. \ref{eq:EI}), and the inversion-symmetric double chain (red, Eq. \ref{eq:ETB}). Note that in the latter case, each of the four bands (eigenvalues of 8$\times$8 Hamiltonian $\mathcal{H}_{\text{TB}}$, Eq. \ref{eq:TB_8x8}) is doubly degenerate. Inset: Corresponding real-space models, with inequivalent $(A,B)$ lattice sites and hopping energies labeled. Note the center of inversion, denoted by a $\ast$, in the double chain (red).\label{fig:TB}}
\end{figure}

A minimal model for a single inversion-asymmetric 1D periodic potential, with a unit cell consisting of unequally-spaced $A$ and $B$ sites, is shown in the blue inset of Fig. \ref{fig:TB}. An appropriate spinless tight-binding Hamiltonian in $k$-space can be written as 
\begin{align}
\mathcal{H}_0=E_0\tau_z+[t_{1}+t_{2}\cos(ka)]\tau_x-t_{2}\sin(ka)\tau_y
\label{eq:H0}
\end{align}
where $t_{1}\neq t_{2}$ are the hopping strengths, $2E_0$ is the difference between on-site energies of atoms $A$ and $B$, and $\tau$s are $2\times 2$ Pauli matrices acting in the $lattice$ sector of Hilbert space.

The spectrum of this single-chain Hamiltonian is given by
\begin{equation}
\mathcal{E}_0(k)^2=E_0^2+t_{1}^2+t_{2}^2+2t_{1}t_{2}\cos(ka),\label{eq:E0}
\end{equation}
and the bandstructure within the first Brillouin zone is shown by the two dashed blue dispersion curves in Fig.~\ref{fig:TB}, using arbitrary parameters $E_0$ and $t_{1,2}$. If including spin, we expand Hilbert space to a $lattice\otimes spin$ basis ($\mathcal{H}_0\otimes \sigma_0$, where $\sigma_0$ is the $2\times 2$ identity), and both these gapped bands acquire trivial twofold spin degeneracy.

The simplest symmetry-allowed lowest-order spin-orbit coupling can be written in this basis as $\mathcal{H}_{\text{SO}}=\alpha (\tau_y \otimes \sigma_z)$, which takes into account the SOC within the unit cell, between the same spins of $A$ and $B$. $\mathcal{H}_{\text{SO}}$ properly commutes with the time reversal operator, $\mathcal{T}=\sigma_yK$, where $K$ is complex conjugation. We note that this form of $\mathcal{H}_{\text{SO}}$ does {\textit{not}} commute with inversion, $\mathcal{I}=\tau_x\otimes\sigma_0$, so $\alpha\neq 0$ only if $E_0\neq 0$. 

The new Hamiltonian is therefore $\mathcal{H}_\text{I} = \mathcal{H}_0\otimes \sigma_0+\mathcal{H}_{\text{SO}},$ which has eigenvalues given by 
\begin{equation}
\mathcal{E}_I(k)^2=\mathcal{E}_0(k)^2+\alpha^2 \pm2\alpha t_2\sin(ka).\label{eq:EI}
\end{equation}
When $\alpha\neq 0$, the last term in Eq. (\ref{eq:EI}) breaks the spin degeneracy and results in four nondegenerate bands everywhere except TRIM points at the zone center and boundary where $\sin (ka)=0$, as shown by the solid blue lines in Fig.~\ref{fig:TB}. The four eigenvectors diagonalizing $\mathcal{H}_\text{I}$ at a general $k$ also diagonalize $\tau_0\otimes\sigma_z $, and thus have definite spin up/down along $z$, due to the form of $\mathcal{H}_{\text{SO}}$. 

Now we will restore inversion symmetry by adding this chain to another, identical to the first except the order of $A$ and $B$ sites are reversed, as shown in the red inset to Fig.~\ref{fig:TB}. The Hamiltonian $\mathcal{H}_\text{II}$ for this second chain is related to the first ($\mathcal{H}_\text{I}$) by unitary transformation with the inversion operator $\mathcal{I}$ and $k\rightarrow -k$, resulting in a sign change to $\mathcal{H}_{\text{SO}}$ and the first term of Eq.~(\ref{eq:H0}), both due to anti-commutativity of $\tau$ operators.

The simplest coupling of these two mutually inverted layers (via hopping between vertically adjacent $A$ and $B$ sites) is given in a $layer \otimes lattice \otimes spin$ basis by $\mathcal{H}_{\text{int}}=t (\lambda_x\otimes \tau_0\otimes \sigma_0)$, where $\lambda$s are $2\times 2$ Pauli matrices in the $layer$ sector of Hilbert space. Our complete Hamiltonian for the double chain is then 
\begin{equation}
\mathcal{H}_{\text{TB}}=\mathcal{H}_\text{I}\oplus \mathcal{H}_\text{II}+\mathcal{H}_{\text{int}},
\label{eq:TB_8x8}
\end{equation}
which is (unlike $\mathcal{H}_{I,II}$) invariant under inversion $\mathcal{I}=\lambda_x\otimes \tau_x \otimes \sigma_0$ (and $k\rightarrow -k$) and satisfies the condition for time-reversal invariance 
$\mathcal{H}_{\text{TB}}(k)=\sigma_y\mathcal{H}^*_{\text{TB}}(-k)\sigma_y$. 
The eigenvalues of $\mathcal{H}_{\text{TB}}$ are given by 
\begin{align}
\mathcal{E}_{\text{TB}}(k)^2&=\mathcal{E}_0(k)^2
+\alpha^2 + t^2 \notag\\ 
&\pm 2\sqrt{\alpha^2 t_2^2\sin^2(ka)+t^2[t_1^2+t_2^2+2t_1t_2\cos(ka)]}.\label{eq:ETB}
\end{align}

Before the interchain hopping is turned on ($t=0$), the energetic spectrum is trivially doubly degenerate due to the block-diagonal Hamiltonian. The eigenvector wavefunctions simultaneously diagonalize both the full spin operator $S_z=\lambda_0 \otimes \tau_0 \otimes \sigma_z$ and the layer measurement operator $\Lambda=\lambda_z \otimes \tau_0 \otimes \sigma_0$; thus they can be chosen to have definite spin {\em and} definite location on one chain or the other. Compared with the `pure' spin-degeneracy of the single chain Hamiltonian ($\mathcal{H}_0\otimes \sigma_0$) when SOC is absent, the double degeneracy here is due to the equivalence (through inversion) of the two chains, even though the two degenerate states residing on opposite chains are indeed opposite in spin. Apparently, the restoration of inversion symmetry \textit{does not} change the reality of the spin-splitting of each individual chain. In other words, bare inversion symmetry without coupling merely adds a copy of the spectrum with opposite spin, rather than removing the original spin-splitting as assumed.\cite{Partoens_NatPhys14} Therefore, if an experimental probe is sensitive only to one of the two chains (or layers) forming a globally centrosymmetric system, the detection of spin-splitting spectrum (other than at the TRIM points) is by no means remarkable, in light of the noncentrosymmetric property of a single chain (or single layer). 

When interlayer hopping is present with $t\neq 0$, the double degeneracy persists, while spin polarization vanishes near the TRIM points. As shown in the red spectrum in Fig.~\ref{fig:TB} and the second term under the square root of Eq.~(\ref{eq:ETB}), gaps develop at the Brillouin zone center and boundaries where the single-chain bands cross. The Hamiltonian still commutes with time reversal and space inversion so we retain twofold degeneracy. However, the eigenvector wavefunctions can no longer be chosen to have definite residence on either chain. Although the full $S_z$ spin operator still commutes with the Hamiltonian, $\Lambda$ does not. Most relevant to claims of `hidden spin polarization', the operator for spin isolated on {\emph {one}} layer $S^{(I)}_z=\text{diag}(1,0) \otimes \tau_0 \otimes \sigma_z$ [or on the other, $S^{(II)}_z=\text{diag}(0,1) \otimes \tau_0 \otimes \sigma_z$] does {\emph{not}} commute with $H_{\text{TB}}$, so eigenstates cannot be classified by definite spin orientation and definite residence on one chain or the other. 

\subsection{Bilayer Bychkov-Rashba}

An alternative to the tight-binding model, where the effect of spin-orbit interaction is built-in by an implicit spatial inversion asymmetry, is the effective 2D Bychkov-Rashba\cite{Bychkov_JETPL84} Hamiltonian that is valid in the vicinity of TRIM points where terms linear in $k$ dominate. In this model, we can use a $layer\otimes spin$ basis to define 
\begin{align}
\mathcal{H}_{\text{BR}}=\alpha \lambda_z\otimes(k_y\sigma_x-k_x\sigma_y)+t (\lambda_x\otimes \sigma_0).\label{eq:BR}
\end{align}

Like the previous 1-D example, this 4$\times$4 Hamiltonian can be analytically diagonalized. Employing a scheme similar to parameterization of two-level systems on the Bloch sphere, we define $d = \sqrt{\alpha^2k^2+t^2}$, $\theta = \arccos(t/d)$ and $\phi = \arctan(k_y/k_x)$.  With basis functions of \{$|{I}{\uparrow}\rangle$, $|{I}{\downarrow}\rangle$, $|{II}{\uparrow}\rangle$, $|{II}{\downarrow}\rangle$\} ($I$ and $II$ are the layer indices), the doubly-degenerate eigenstates are superpositions of
\begin{align}
\psi_1 = \frac{1}{\sqrt{2}}\begin{bmatrix}
i\sin\frac{\theta}{2} \\
\nu e^{i\phi}\cos\frac{\theta}{2}\\
-\nu i\sin\frac{\theta}{2} \\
e^{i\phi}\cos\frac{\theta}{2}
\end{bmatrix},\,\,
\psi_{2} = \frac{1}{\sqrt{2}}\begin{bmatrix}
\nu i\cos\frac{\theta}{2} \\
e^{i\phi}\sin\frac{\theta}{2}\\
i\cos\frac{\theta}{2} \\
-\nu e^{i\phi}\sin\frac{\theta}{2}
\end{bmatrix},\label{eq:BRstates}
\end{align}
in which $\nu = \pm 1$ corresponds to the eigenvalues $\pm d$. 

With vanishing coupling $t$ between the two layers, $\mathcal{H}_{\text{BR}}$ is $2\times 2$ block-diagonalized, so the layer index is a good quantum number and the wavefunctions can be fully projected onto the local layer basis with projection operators $\hat{P}_{I}=|{I}{\uparrow}\rangle\langle{I}{\uparrow}|+|{I}{\downarrow}\rangle\langle{I}{\downarrow}|$, etc. This situation corresponds to $\theta = \pi/2$; by taking the sum and difference of the two states in Eq.~(\ref{eq:BRstates}),  the nonzero wavefunction components are reduced to the familiar form $(\pm i,e^{i\phi})^\text{T}$ of the single-layer Bychkov-Rashba Hamiltonian eigenstates. Again, bare restoration of inversion symmetry without coupling does not change the spin texture of a single layer in $k$-space. Using the `projected' spin operator $\bm{S}_{I,II} = \hat{P}_{I,II}^\dagger (\lambda_0\otimes\bm{\sigma}) \hat{P}_{I,II}$, we see that $\bm{S}_{I,II}\cdot \bm{n}$ commutes with $\mathcal{H}_{\text{BR}}$, where the unit vector $\bm{n}=\pm(\sin\phi,-\cos\phi,0)^\text{T}$ points along the Bychkov-Rashba field.

Nonvanishing $t$ couples states localized on opposite layers and opens a gap at the TRIM point $|k|=0$. Now that $\theta\neq \pi/2$, we see from Eq.~(\ref{eq:BRstates}) that there is no way of forming a suitable superposition having nonzero components only on one layer; the nonzero off-diagonal blocks of $\mathcal{H}_{\text{BR}}$ prevent its commutativity with $\bm{S}_{I,II}\cdot \bm{n}$. Nonetheless, using the two states in Eq.~(\ref{eq:BRstates}), one can still calculate the expectation values
\begin{align}
\langle\bm{S}_{I,II}\rangle_{1,2} = \frac{1}{2}
\begin{bmatrix}
\nu\mu\sin\phi\sin\theta \\
-\nu\mu\cos\phi\sin\theta \\
\eta\cos\theta
\end{bmatrix}.\label{eq:BRsigma}
\end{align}
Here, $\mu=1 (-1)$ corresponds to projection on layer~$I$ ($II$), and $\eta=1(-1)$ to one of the two degenerate states $\psi_2$ ($\psi_1$). Summation over the state index $\eta$ results in the `hidden spin polarization' $\pm\sin\theta(\sin\phi,-\cos\phi,0)^\text{T}$. Compared with the $t=0$ case, the amplitude of the local Bychkov-Rashba spin texture is now reduced by $\sin\theta=\alpha |k|/d$, which vanishes as $k$ approaches the TRIM point and $\theta\rightarrow 0$ or $\pi$ (the Bloch sphere poles). 

This is the expectation for a spin-resolved and layer-selective probe. However, with a (hypothetical) detection scheme that can directly project out one of the $\psi_1$ or $\psi_2$ basis states, Eq.~(\ref{eq:BRsigma}) astonishingly predicts a finite out-of-plane spin component maximized at the TRIM point, despite the fact that Eq.~(\ref{eq:BR}) has no $\sigma_z$ term! It is clear that this unexpectedly nonzero value is merely a vestige obtained by the particular (and ultimately arbitrary) way Eq.~(\ref{eq:BRstates}) segregates wavefunction components into the chosen basis and is subject to the choice of gauge.

\subsection{`Sectors' with inversion symmetry}

In both models discussed above, each individual `sector' is inversion asymmetric with intrinsically spin-split bands. This situation differs from the 3D diamond structure mentioned in the introduction, where each FCC sublattice `sector' is inversion symmetric. Yet, the two complementary sectors are indeed distinct in the sense that periodic translation does not transform one into the other. This type of lattice is different from other inversion symmetric systems that cannot be decomposed into complementary `sectors' unrelated by full translation operations, such as the simple cubic lattice that we will focus on in the following section. At the end of this section however, we take the diamond structure as an example and analyze  the finite and opposite spin operator expectation values of the two `sector' projections of the wavefunctions.

It is well known that an external electric field $\bm{E}$ can induce spin-splitting in an otherwise inversion-symmetric band structure at general $\bm{k}$.\cite{Winkler_book} This phenomenon can readily be understood via third-order $\bm{k}\cdot \bm{p}$ perturbation theory, due to the polar vector nature of both the electric field potential $e\bm{E}\cdot{\bm{r}}$ and the $\bm{k}\cdot\bm{p}$ term. For example,  $s$-like bands with scalar symmetry can couple to polar vector $p$-like bands via the position operator $\hat{\bm{r}}$; after the axial vector spin-orbit interaction mixes the $p_{x,y,z}$-like bands, the perturbation pathway ends by coupling back to the initial $s$-like bands via the $\bm{k}\cdot\bm{p}$ term. The net effect is an extra term $\propto\bm{E}\times \bm{k}\cdot \bm{\sigma}$ in the $s$-band effective Hamiltonian, leading to a lowest order splitting linear in $\bm{E}$ and $\bm{k}$ between opposite and otherwise degenerate spin-eigenstates. 

In the example of a FCC sublattice in the diamond structure, it is the crystal potential of the complementary sector that provides such an `external electric field', which mediates the intrinsic spin-orbit coupling together with the crystal momentum of the electronic wavefunctions to induce the spin-splitting of the band structure of an individual FCC sector. The resulting eigenstates at general $\bm{k}$ have nonzero expectation value of spin angular momentum. In the same manner, the complementary sector experiences an `external electric field' in the opposite direction so its own spin-splitting amplitude is reversed in sign. Therefore, combining band structures of both sectors restores global double-degeneracy, and the crystal fields further open gaps at TRIM points of the full band structure in the same way the nonvanishing $t$ does in Eq.~(\ref{eq:ETB}). 

This `equivalent external electric field' $\bm{E}$ provided by the complementary sector is, in general, $\bm{k}$-dependent. For wavevectors on certain high symmetric axes like [100] and [111], $\bm{E}$ and $\bm{k}$ are aligned, and $\bm{E}\times \bm{k}\cdot \bm{\sigma}$ vanishes. For the electronic states of the full diamond structure along these high symmetric axes, projection onto a given 'sector' results in zero expectation value of the spin angular momentum.

As pointed out in Ref.[\onlinecite{Zhang_NatPhys14}], this scenario only happens for certain inversion symmetric lattices in which the local potential at each lattice site lacks inversion symmetry. Our $\bm{k}\cdot\bm{p}$ analysis makes clear that, by summing over the complementary sector lattice, such `local inversion asymmetry' guarantees a nonzero matrix element of $\bm{E}\cdot{\bm{r}}$ between the $s$- and $p$-like states, which are Bloch wavefunctions extending throughout the entire lattice. However, the only role played by this equivalent `external electric field' is in completing the spin-splitting perturbation pathway which includes SOC that the FCC sector lattice already possesses. Therefore, it is incorrect and misleading to make assertions like ``spin-orbit coupling effects are governed by the local symmetry of the potential felt by the electron, rather than by the symmetry of the bulk crystal''.\cite{Cottin_NatComm16}  Ironically, it has been known for many years that the origin of intrinsic spin-orbit matrix elements is actually even much more `local' than suggested, mainly arising from the part of the wavefunctions orthogonal to $p$-like core electrons in the inner shell region,\cite{Liu_PR62,Saravia_PR68,Chelikowsky_PRB76} where the electrostatic potential varies most drastically.

\section{generic surface spin splitting}

The surface of {\emph{any}} lattice fundamentally breaks inversion symmetry, regardless of whether the semi-infinite bulk underneath is centrosymmetric. In this sense, any Bloch band structure at the surface should be spin-split for general $\bm{k}$. Such spin-splitting can be pronounced, and is often complicated by surface reconstruction. Therefore, the attempt to make experimental contact to the notion of `hidden spin polarization' using surface sensitive techniques like ARPES \cite{Riley_NatPhys14,Bertoni_PRL16, Razzoli_PRL17, Yao_NComm17} is inevitably complicated.\cite{Armitage_RMP18}

\begin{figure}
\includegraphics[width=3.75in, trim={0.8cm 0.2cm 0 0},clip]{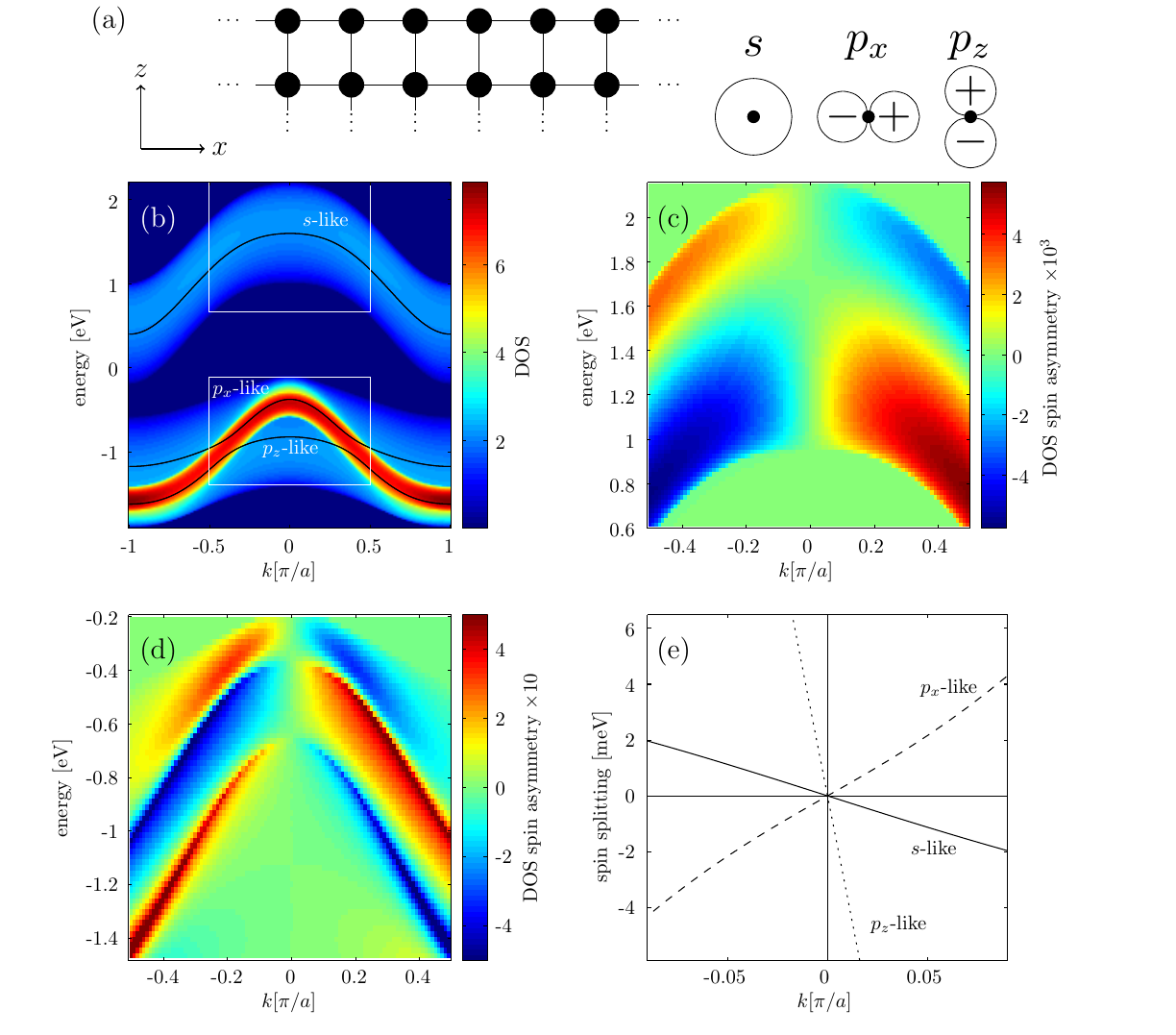}
\caption{(a) 2D semi-infinite half-space square lattice with surface normal to $z$, and $s,p_x,p_z$ orbital basis per site. (b) Surface density of states, with unbroadened (and spin-degenerate) bands of an isolated 1D wire overlain in black. (c),(d) Difference in spin-dependent DOS, induced by symmetry breaking at the surface, near $k_x=0$ for $s$-like bands and $p$-like bands, respectively, within regions denoted by white boxes in (b). (e) $k_x$-dependent spin splitting energy for all three bands. Parameters used: $E_s=1$eV,
$E_p=-1$eV, $V_{pp\pi}=0.1$eV, $V_{ss}=V_{pp\sigma}=V_{sp}=0.3$eV, on-site spin-orbit energy $\Delta=0.1$eV, and numerical broadening $\eta=10^{-3}$eV. \label{fig:surface}}
\end{figure}

To illustrate this point, we implement a tight-binding model on the centrosymmetric simple square lattice in the $x$-$z$ plane with lattice constant $a$, using a $\{s,p_x,p_z\}$ atomic orbital basis shown in Fig. \ref{fig:surface}(a) including on-site SOI. This system can be constructed by coupling isolated 1D wires oriented along $x$, each with a Bloch Hamiltonian $H_{\text{wire}}=$
\begin{align}
\resizebox{0.9\hsize}{!}{
$\begin{bmatrix}
E_s+2V_{ss}\cos k_xa & 2iV_{sp}\sin k_xa &0\\
-2iV_{sp}\sin k_xa & E_p+2V_{pp\sigma}\cos k_xa& 0 \\
0& 0 & E_p+2V_{pp\pi} \cos k_xa
\end{bmatrix}$,}
\label{eq:wireH}
\end{align}
with $E_s$ and $E_p$ the onsite energies of the $s$ and $p_{x,z}$ orbitals, respectively. $V_{ss}$, $V_{sp}$, $V_{pp\pi}$ and $V_{pp\sigma}$ are coupling integrals between orbitals.\cite{Slater_PR54} SOC is accounted for by expanding to a $lattice \otimes spin$ basis and including an additional matrix element  $i\Delta \sigma_y$ coupling opposite spins of $p_x$ and $p_z$, which emerges from the axial vector symmetry of the SOI operator. The coupling matrix linking this wire to its nearest neighbor along $z$ is then simply
\begin{align}
t_z=
\begin{bmatrix}
V_{ss}&0&V_{sp}\\
0&V_{pp\pi}&0\\
-V_{sp}& 0& V_{pp\sigma}
\end{bmatrix}.
\label{eq:wiret}
\end{align}
The non-Hermitian property of this matrix, due to opposite parity of the $s$ and $p_z$ orbitals upon mirror reflection in the $z$ direction, is essential in what follows.

Spin splitting at the abrupt surface can be understood to arise (for the $s$-like band) from matrix elements coupling the surface layer $s$ orbital to $p_x$ in the translationally-invariant direction (off-diagonal in Eq. \ref{eq:wireH}, a $k_x$-dependent term), on-site SOI which couples $p_x$ to $p_z$ (a $\sigma_y$ term), then vertical hopping to the $s$ and $p_z$ orbitals in the layer below it (off-diagonal or $V_{pp\sigma}$-dependent, respectively, in Eq. \ref{eq:wiret}), and back again to the initial $s$ orbital in the surface layer, with an energy denominator $\sim E_g^2$ in this third order perturbation ($E_g=E_s+2V_{ss}-E_p-2V_{pp(\pi,\sigma)}$ is the direct bandgap between the $s$- and $p_{(z,x)}$-like bands, respectively). Similar perturbation paths apply to $p$-dominated bands, although the energy denominator in this case is $\sim 2E_g(V_{pp\sigma}-V_{pp\pi})$ resulting in stronger spin splitting. Importantly, one of the last two steps involves coupling between the odd-parity $p_z$ and the even $s$ orbital on different layers, playing the role of an electric field normal to the surface. This coupling induces an effective interaction $\propto k_x\sigma_y$ that would otherwise be canceled by coupling to the complementary half-space in an infinite bulk.

Using Eqs. \ref{eq:wireH} and \ref{eq:wiret}, we can calculate the density of states (DOS) at the 1D surface (parallel to $x$, normal to $z$). First, the Green's function is found by solving Dyson's equation\cite{Dy_PRB79, Bravi_PRB14}
\begin{align}
g=&[ (E+i\eta)I-H(k)-t_zgt_z^\dagger ]^{-1},
\end{align}
where $H(k)$ includes SOI as described above, and $\eta$ is numerical broadening\cite{Datta_B95}. Then, the DOS is computed via $D(E,k)=\frac{1}{\pi}\Im \text{Tr} g$. 

Fig.~\ref{fig:surface}(b) shows the electronic structure within the full surface BZ, consisting of broadened 1D bands of the isolated 1D wire (black curves). For convenience we call the three bands $s$-, $p_x$- and $p_z$-like, based on their mostly pure orbital nature close to the zone center. SOC is evident in the avoided crossing between $p_x$- and $p_z$-like bands at lower energy around $k_x = 0.5 \pi/a$. Peak DOS for the highly dispersing $\sigma$-bonded $p_x$-like band is greater than the others because it is broadened less by weak $\pi$-bond coupling to the semi-infinite bulk. 

Spin splitting is not immediately apparent in the total DOS. Therefore, we use the difference between spin-dependent DOS, obtained by partial trace of the Green's function, to find the relative spin polarization oriented along $y$ within the regions enclosed by white boxes in Fig. \ref{fig:surface}(b). The region enclosing the $s$-like band is shown in Fig. \ref{fig:surface}(c), and the $p$-like bands are shown in Fig. \ref{fig:surface}(d). Spin polarization in each has odd parity across $k_x=0$, reflecting time-reversal symmetry. On top of the larger spin-orbit coupling strength of the $p_x$-like states due to the smaller energy denominator of the perturbation paths, their narrower broadening further enhances their spin polarization compared with that of the $s$-like states. By locating the peak maxima in spin-dependent DOS, we can numerically calculate the linear $k$-dependent spin splitting for each band, as shown in Fig. \ref{fig:surface}(e). Although it is much greater for the $p$-like bands, even the $s$-like band spin splitting is significantly nonzero.  

Note that upon extension to a 3D simple cubic lattice (which is centrosymmetric and cannot be decomposed into complementary `sectors' related only by inversion) with a surface in the $x$-$y$ plane, this effect has the familiar Bychkov-Rashba form $\propto k_x\sigma_y-k_y\sigma_x$. As we have shown, if the bulk electronic states are projected onto a given surface, the expectation value of spin-angular momentum in a given band at generic $\bm{k}$ is nonzero; its degenerate partner is projected on the opposite parallel surface and has opposite spin. However, the associated spin operator does not commute with the total Hamiltonian of the lattice and thus does not reflect a conserved quantity of the bulk. 

\section{Discussion \& Conclusion}

We have examined the notion of `hidden spin polarization' from simple tight-binding and effective-Hamiltonian perspectives (neither of which rely on sophisticated but opaque computational methods like density functional theory). Although the decomposition procedure suggested in Refs.~[\onlinecite{Zhang_NatPhys14,Liu_PRB15}] is {\textit{mathematically}} valid, interpreting its subsequent implications requires several important caveats that must be emphasized.

First, true spin polarization of Bloch bands must be a consequence of Hamiltonian commutativity with a unique spin operator. In both cases, finite coupling with $t\neq 0$ between constituent sublattices makes it impossible to choose a layer-dependent spin basis, especially when $k$ is close to TRIM points where eigenstate localization purity is suppressed.

Second, preferential projection of wavefunction components from one sublattice implicitly entails a choice of basis within the twofold degenerate subspace of each orbital Bloch band. Of course, in general one is free to represent degenerate states in any convenient arbitrary superposition; the expectation value of any operator summed over the full subspace is invariant to this choice. However, expectation values of a {\emph{single}} state selected from this degenerate subspace critically depend on the choice of basis. Outcomes of such biased measurements therefore say as much about the geometry and mechanisms employed in the experiment than intrinsic electronic structure {\textit{per se}}.

This is particularly true for experiments designed to support the notion of `hidden spin polarization'. Methods selectively probing a single exposed surface, e.g. via ARPES\cite{Riley_NatPhys14,Bertoni_PRL16, Razzoli_PRL17, Yao_NComm17}, inherently perform a biased projection onto a spatially localized basis. Spin splitting observed in the empirical data must be interpreted in the context of this broken spatial symmetry, and not solely as a probe of bulk states.

Finally, we note that the opposite expectation values of the spin operator evaluated from the sublattice projections of delocalized Bloch electronic states resembles the alternating orientation of local magnetic moments as in antiferromagnetic order. However, as mentioned in Ref.~[\onlinecite{Jungwirth_NatNano16}], even when a momentum imbalance is created by an external electric field, ``In Si, there is no equilibrium antiferromagnetic order that can be manipulated by these local staggered non-equilibrium polarizations''. Antiferromagnetism spontaneously breaks spatial symmetries of the lattice sites and time-reversal symmetry, so Kramers' theorem does not apply; a complete description of the symmetry invokes the magnetic space group\cite{Tinkham_book}. Moreover, magnetically ordered phases arise below a transition temperature due to (exchange) \textit{interactions} in a \textit{multiparticle} system, whereas bandstructure is fundamentally a single-particle description of the energetic spectrum.
Given these essential dissimilarities, it is misleading to equate the concepts discussed in the present manuscript with true antiferromagnetism.
% The analogy of such disparate systems.
%It is simply inappropriate to conflate such disparate systems. 

We genuinely hope that our detailed discussion makes clear what `hidden spin polarization' is (and is not).  

\begin{acknowledgments}
We acknowledge support from 
the Defense Threat Reduction Agency under contract HDTRA1-13-1-0013 and the National Science Foundation under contract ECCS-1707415. 
\end{acknowledgments}

\bibliography{Kramers}

\end{document}